\documentclass[prl,twocolumn,superscriptaddress,showpacs]{revtex4-1}
\usepackage{amsmath}
\usepackage{amsfonts}
\usepackage{graphicx}
\usepackage{pst-all}
\usepackage{graphics}
\usepackage{amssymb}
\usepackage{layout}
\usepackage{verbatim}
\usepackage{amsfonts,epsfig}
\usepackage{bm}
\usepackage{amsmath}

\newcommand{\ket}[1]{|#1\rangle}

\begin{document}

\title{Topological quantum buses: coherent quantum information transfer between topological and conventional qubits}
\author{Parsa Bonderson}
\author{Roman M. Lutchyn}
\affiliation{Microsoft Research, Station Q, Elings Hall, University of California, Santa Barbara, CA 93106, USA}
\date{\today}

\begin{abstract}
We propose computing bus devices that enable quantum information to be coherently transferred between topological and conventional qubits. We describe a concrete realization of such a topological quantum bus acting between a topological qubit in a Majorana wire network and a conventional semiconductor double quantum dot qubit. Specifically, this device measures the joint (fermion) parity of these two different qubits by using the Aharonov-Casher effect in conjunction with an ancilliary superconducting flux qubit that facilitates the measurement. Such a parity measurement, together with the ability to apply Hadamard gates to the two qubits, allows one to produce states in which the topological and conventional qubits are maximally entangled and to teleport quantum states between the topological and conventional quantum systems.
\end{abstract}

\pacs{03.67.Lx, 03.67.Pp, 71.10.Pm, 05.30.Pr}
\maketitle









%

{\it Introduction.}---The topological approach to quantum information processing obtains its exceptional fault-tolerance by encoding and manipulating information in non-local (topological) degrees of freedom of topologically ordered systems~\cite{Kitaev97,Freedman98,Nayak08}. These non-local degrees of freedom do not couple to local operations, so the error rates for topological qubits and computational gates are exponentially suppressed with distance between anyons and inverse temperature, providing an enormous advantage over conventional quantum computing platforms. However, the same fact also makes it very challenging to coherently transfer quantum information into and out of topological systems, since this not only requires coupling the non-local degrees of freedom in the topological system to an external system, but doing so in a controlled and coherent manner. In other words, one must be able to create quantum entanglement between the topological and conventional states. In this letter, we propose a device that can entangle and coherently transfer quantum information between topological and conventional quantum media, i.e. a ``topological quantum bus.'' Such devices would allow one to harness the relative strengths of the different quantum media and could prove crucial for the implementation of quantum computation.

A prime example of how topological quantum buses would be useful stems from the fact that a computationally universal gate set cannot be produced for Ising anyons using topologically-protected braiding operations alone. Unless one has a truly topologically ordered Ising system (which is not the case for superconductor-based systems, including Majorana wires) and can perform certain topology changing operations~\cite{Bravyi00-unpublished,Freedman06,Bonderson10}, one will need to supplement braiding operations with topologically unprotected operations. Fortunately, these can be error-corrected for a high error-rate threshold of approximately $0.14$ by using the topologically-protected Ising braiding gates to perform ``magic-state distillation''~\cite{Bravyi05}. Within a topological system, one can generate unprotected gates, for example by bringing non-Abelian anyons close to each other, which generically splits the energy degeneracy of the topological state space~\cite{Bonderson09b} and hence dynamically gives rise to relative phases, or by using interfering currents of anyons~\cite{Bonderson'10}, which can have an equivalent effect. However, besting even such a high error threshold may still prove difficult using unprotected operations within a topological system, as a result of significant non-universal effects. A topological quantum bus would allow one to import the necessary topologically unprotected gates from conventional quantum systems, for which error rates below $0.14$ have already been achieved~\cite{Ladd10}.

A robust method of implementing quantum buses is through the use of measurements in an entangled basis, e.g. Bell state measurements. For a topological quantum bus, this can be achieved by a measurement of the joint parity of a topological-conventional qubit pair, given the ability to perform Hadamard gates on any qubit (as we explain later in detail).
Joint parity measurements corresponds to the two orthogonal projectors
\begin{eqnarray}
\label{eq:Pi_0}
\Pi_{0} &=& \left| 00 \right\rangle \left\langle 00 \right| + \left| 1 1 \right\rangle \left\langle 11 \right|, \\
\label{eq:Pi_1}
\Pi_{1} &=& \left| 01 \right\rangle \left\langle 01 \right| + \left| 10 \right\rangle \left\langle 10 \right|,
\end{eqnarray}
where $\left|0\right\rangle$ and $\left|1\right\rangle$ are the logical basis states of the qubits. Topological systems, however, tend to be rather obstructive to such hybridization with external systems.
For example, quantum Hall states (the archetypal topological systems) require a large background magnetic field, which ``destroys" superconductivity and eliminates the possibility of coupling to Josephson-junction qubits.

Fortunately, the recently proposed implementations~\cite{Lutchyn'10,Oreg'10} of Majorana nanowires~\cite{Kitaev'01} appear promising for overcoming such obstacles. These wires localize zero energy Majorana fermions at their endpoints and as such provide a one-dimensional topologically protected two-level system. At first, this may seem too simple a system, providing a topological qubit, but lacking quantum information processing. However, one can form a network of Majorana wires and manipulate them using gate electrodes~\cite{Hassler'10, Aliceaetal'10} in a manner that performs braiding exchanges of their endpoints, and hence their respective Majorana fermions. Remarkably, this generates the topologically-protected braiding operations of Ising anyons (up to an overall phase) on the topological state space~\cite{Aliceaetal'10}. It follows that Majorana wire networks can be utilized as Ising anyons for topologically-protected quantum information processing. In this letter, we provide a concrete realization of a topological quantum bus that uses the Aharonov-Casher effect~\cite{Aharonov84} to coherently transfer quantum information between a topological qubit in a Majorana wire system and a conventional semiconductor double-dot qubit~\cite{charge_qubit'03,Hanson_RMP'07}.

The Aharonov-Casher effect is dual to the more familiar Aharonov-Bohm effect and involves interference of particles with magnetic moment (vortices) moving around a line charge. It enables one to perform non-local measurements of charge in a region by utilizing the interference of vortices traveling through two different paths around the region. For superconducting systems it is natural to try to use Abrikosov vortices in this context~\cite{Akhmerov'09,Fu'09,Grosfeld'10,Sau10b}. However, Abrikosov vortices in most $s$-wave superconductors have rather large mass due to the large number of subgap states localized in their cores~\cite{Volovik'97} and, as a result, these vortices behave classically. An alternative is to use Josephson vortices (fluxons), which arise due to phase-slip events in Josephson junctions~\cite{Friedman_prl'02,Tiwari_prb'07}. Their effective mass is determined by the charging and Josephson energies of the junction and can be much smaller than that of Abrikosov vortices, allowing them to behave quantum-mechanically~\cite{Elion'93,Wallraff'03}. Indeed, the Aharonov-Casher effect with Josephson vortices has been experimentally observed~\cite{Elion'93} and several proposals have been made to utilize it in the context of topological quantum information processing~\cite{Hassler'10,Nilsson'10,Sau_network,Clarke'10}.

\begin{figure}
\includegraphics[width=3.0in,angle=0]{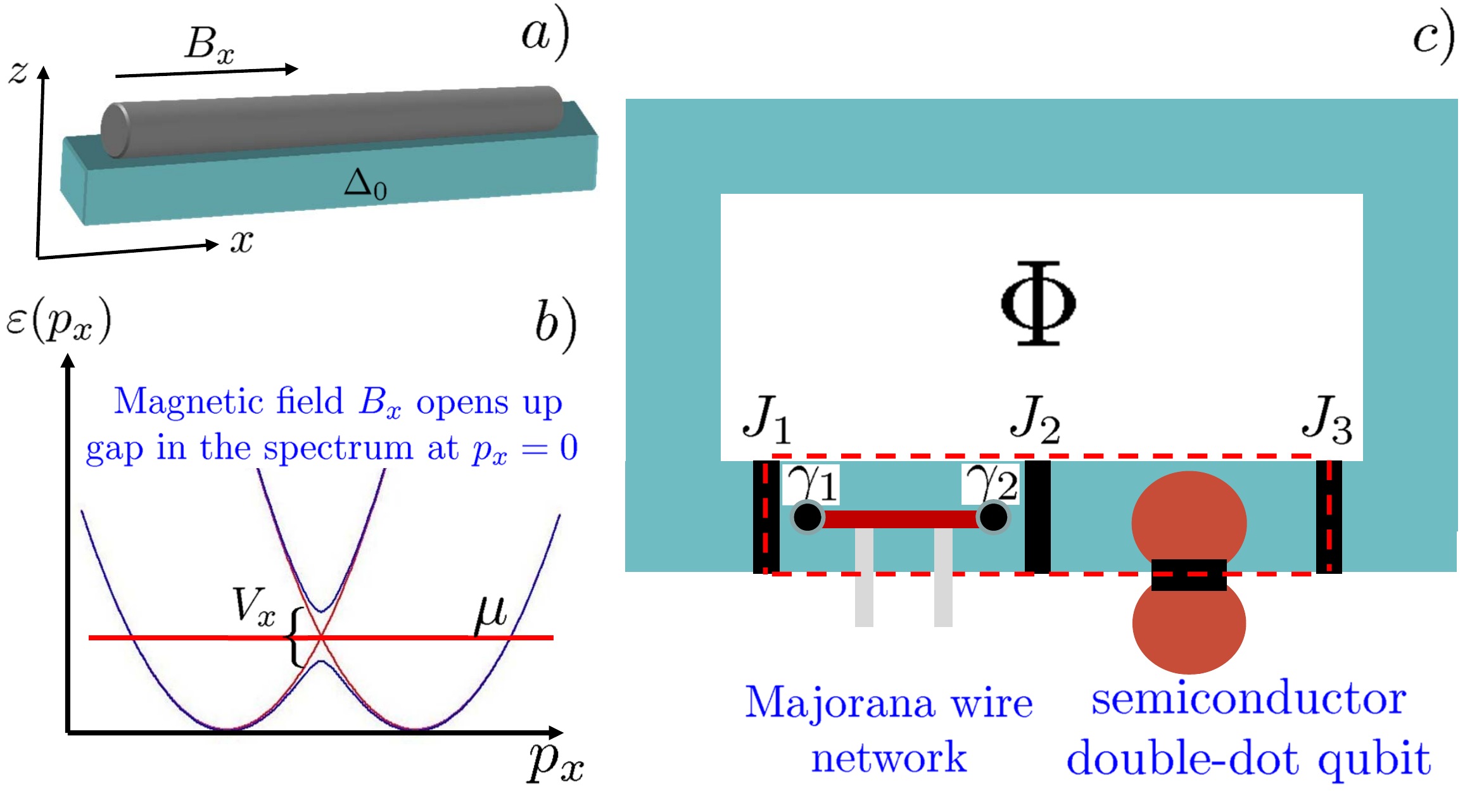}
\caption{(Color online) a) A semiconductor nanowire coupled by proximity with an $s$-wave superconductor, in the presence of an in-plane magnetic field $B_x$. b) The energy dispersion for the semiconductor with spin-orbit coupling in a magnetic field, which opens the gap in the spectrum. When the chemical potential $\mu$ is in this gap, the nanowire coupled with an $s$-wave superconductor is driven into the topological phase. c) A schematic of the proposed device used to entangle topological and conventional qubits (e.g. a Majorana wire qubit and semiconductor double-dot qubit). A flux qubit consisting of three Josephson junctions (the black strips labeled $J_1$, $J_2$, and $J_3$) supports clockwise or counter-clockwise supercurrent. When $E_{J_1}=E_{J_3}$, there is interference between quantum phase slips across junctions 1 and 3. These phase slips correspond to Josephson vortex tunneling encircling the superconducting islands as shown by the dashed (red) line. Via the Aharonov-Casher effect, quantum interference of vortices around the islands produces an energy splitting for the flux qubit (at its degeneracy point) that strongly depends on the state of the topological and conventional semiconductor qubits. The topological and non-topological segments of the nanowires are represented by red and grey, respectively. The latter can be achieved by driving the wire into the insulating or trivial superconducting phases.}
\label{Fig1}
\end{figure}

{\it Topological qubit.}---The basic element in the proposed implementation of Majorana wires of Refs.~\cite{Lutchyn'10,Oreg'10} is a semiconductor nanowire with strong spin-orbit interactions, coupled with an s-wave superconductor, see Fig.~\ref{Fig1}a. The Hamiltonian (with $\hbar=1$) for such a nanowire is:
\begin{eqnarray}
\mathcal{H}_0 &=& \!\! \int_{-L/2}^{L/2} d x \, \psi_{\sigma}^\dag(x) \left(- \frac{\partial_x^2}{2m^*} - \mu \right. \notag \\
&& \qquad \qquad \,\, \left. \phantom{\frac{\partial_x^2}{2m^*}} + i \alpha \sigma_y \partial_x + V_x \sigma_x \right)_{\!\! \sigma \sigma'} \psi_{\sigma'}(x),
\label{eq:H0}
\end{eqnarray}
where $m^*$, $\mu$, and $\alpha$ are the effective mass, chemical potential, and strength of spin-orbit Rashba interaction, respectively, and $L$ is the length of the wire, which is taken to be much longer than the effective superconducting coherence length $\xi$ in the semiconductor. An in-plane magnetic field $B_x$ leads to spin splitting $V_x\!=\!g_{\rm SM}\mu_B B_x/2$, see Fig.~\ref{Fig1}b, where $g_{\rm SM}$ is the $g$-factor in the semiconductor and $\mu_B$ is the Bohr magneton. When coupled with an $s$-wave superconductor, the nanowire can be driven into a non-trivial topological phase with Majorana zero-energy states localized at the ends. This happens when the chemical potential is properly adjusted and lies in the gap, see Fig.~\ref{Fig1}b. In the simplest case of a single-channel nanowire, the topological phase corresponds to $|V_x|>\sqrt{\mu^2+\Delta^2}$ where $\Delta$ is the proximity-induced pairing potential (see also Ref.~\cite{lutchyn_multi} for the multi-channel case). The two Majorana fermions $\gamma_1$ and $\gamma_2$ residing at the ends of a wire constitute a topological qubit, since they give rise to a two-level system that is degenerate up to $O(e^{-L/\xi})$ corrections that are exponentially suppressed with the length of the wire. Indeed, one can formally define a non-local Dirac fermion operator as $c=\gamma_1+i\gamma_2$ and then the two logical states of the qubit correspond to state in which this Dirac fermion is unoccupied $\ket{0} \equiv \ket{n_p\!=\!0}$ and occupied $\ket{1} \equiv \ket{n_p\!=\!1}$, where $c\ket{n_p\!=\!1}=\ket{n_p\!=\!0}$, $c\ket{n_p\!=\!0}\!=\!0$, and $c^{\dagger}c\ket{n_p}\!= n_p\ket{n_p}$. Thus, the topological qubit states are characterized by fermion parity $n_p\!=\!0,1$. As previously mentioned, in a network of such wires, these Majorana fermions behave as Ising non-Abelian anyons when they are translocated, e.g. using gate electrodes.

{\it Qubit entanglement.}---We now discuss how one can entangle topological and conventional qubits by measuring the fermion parity on the superconducting island using the Aharonov-Casher effect~\cite{Hassler'10}. Consider the superconducting flux qubit with Josephson junctions designed to have left-right symmetry such that Josephson coupling energies $E_{J_1}=E_{J_3}\equiv E_{J}$, see Fig.~\ref{Fig1}c. The two current-carrying states,  clockwise $\ket{\!\circlearrowright}$ and counter-clockwise $\ket{\!\circlearrowleft}$, form the basis states of the flux qubit. When the applied external flux piercing the flux qubit is equal to a half flux quantum, i.e. $\Phi=h/4e$, there is a degeneracy between the two current-carrying states. This degeneracy is lifted by the macroscopic quantum tunneling between the states $\ket{\!\circlearrowright}$ and $\ket{\!\circlearrowleft}$ due to the presence of a finite charging energy of the islands, which tends to delocalize the phase. Thus, the new eigenstates of the qubit are $\ket{\pm}\!=\!(\ket{\!\circlearrowright}\!\pm\! \ket{\!\circlearrowleft})/\sqrt{2}$. For the device shown in Fig.~\ref{Fig1}c the energy splitting between states $\ket{\pm}$ depends on the quantum interference of the fluxon trajectories. Indeed, the total Josephson energy of the qubit is~\cite{flux}
\begin{equation}
\frac{U_J}{E_J}\!=\!-\!\left[\!\cos \varphi_1\!+\!\cos \varphi_2 \!+\! \frac{E_{J_2}}{E_J}\cos\left(\!2\pi \frac{\Phi}{\Phi_0} \!-\! \varphi_1 \!-\! \varphi_2\right) \right],
\end{equation}
where we assume $E_{J_2}>E_{J}$, in contrast with values typically used for flux qubits. The potential $U_J$ reaches its minima at two inequivalent points $(\varphi_1,\varphi_2)=(\pm \varphi^* +2\pi m, \mp \varphi^*\mp 2\pi n)$ for a given $n$ and $m$ which correspond to clockwise and counter-clockwise circulating currents, and $\varphi^*=\cos^{-1}(E_J/2E_{J_2})$. Starting, for example, from the configuration with $(\varphi^*,-\varphi^*)$, there are two paths to tunnel to a different flux state: $(\varphi^*,-\varphi^*)\rightarrow(-\varphi^*+2\pi,\varphi^*)$  and $(\varphi^*,-\varphi^*)\rightarrow(-\varphi^*,\varphi^*+2\pi)$ which correspond to a phase slip through junction 1 or 3, respectively. As a result, there is an interference between the two paths that encircle the middle islands in Fig.~\ref{Fig1}c. (Note that the amplitude for the phase slips across the middle junction is suppressed in this setup since $E_{J_2}>E_J$.) This interference is sensitive to the total charge enclosed by the paths, i.e. the charge residing on the two islands, and is determined by the Berry phase contribution. For the device considered here, the splitting energy is given by $\Delta\!=\!\Delta_0\cos(\phi_{\rm AC}/2)$ where $\phi_{\rm AC}=\pi q/e$ is the Aharonov-Casher phase for total charge on the islands given by $q= en_p + q_{\rm ext}$, where $n_p$ is the fermion occupation of the Majorana wire and $q_{\rm ext}$ is the induced gate charge on the islands~\cite{Hassler'10}. Given that the qubit splitting energy now depends on the fermion occupation number, the state of a topological qubit can be efficiently read out using, for example, the rf reflectometry technique~\cite{Paila}, which can be carried out with sub-microsecond resolution times. It is implicitly assumed here that superconducting islands have the same charging energy yielding the same tunneling amplitude $\Delta_0$. Assuming $E_J/E_c \approx 10$ and $E_{J_2}/E_J \approx 1.25$, WKB approximation gives $\Delta_0 \approx 0.02 h \nu_a$~\cite{flux}, where $\nu_a$ is the attempt frequency, which we estimate to be $\nu_a \sim 0.1-1$~GHz.

We now consider a situation where $q_{\rm ext}$ has a quantum component corresponding to coherent electron tunneling inside the area enclosed by the vortex circulation. This can be realized, for example, by coupling the flux qubit to a semiconductor double quantum dot (DQD) qubit~\cite{Hanson_RMP'07} as shown in Fig.\ref{Fig1}c. We assume here that there is a galvanic isolation between the superconductor and semiconductor, so that there is no charge transfer between them. Remarkably, one can realize DQD qubits using InAs nanowires~\cite{Delft_nanowire}, which may thus serve as a dual-purpose component (also being used for the Majorana nanowires) and reduce the technical challenges of implementing our proposal. If there is a single electron in the DQD, we can define the logical qubit basis states to be: $\ket{0} \equiv \ket{0}_{\rm U} \otimes \ket{1}_{\rm L}$, where the electron occupies the lower quantum dot, and $\ket{1} \equiv \ket{1}_{\rm U}\otimes \ket{0}_{\rm L}$, where the upper quantum dot is occupied, see Fig.\ref{Fig1}c. This situation corresponds to a semiconductor charge qubit~\cite{charge_qubit'03,charge_qubit'09}. If there are two electrons in the DQD, then one can define the logical qubit basis states to be $\ket{0} \equiv \ket{0}_{\rm U} \otimes \ket{2}_{\rm L}$ and $\ket{1} \equiv \ket{1}_{\rm U}\otimes \ket{1}_{\rm L}$, where the electron spins are in the singlet and triplet states, respectively. This situation corresponds to the semiconductor spin qubit~\cite{Hanson_RMP'07}. Both these qubits share one common feature which can be exploited for our purposes: The qubit basis states correspond to the electron parity on the upper dot enclosed by the vortex circulation. If the evolution of the semiconductor qubit is much slower than the measurement time and fluxon tunneling rate, then one can use the flux qubit to entangle topological and conventional qubits via the Aharonov-Casher effect. Indeed, the flux qubit splitting energy $\Delta$ is the same for combined topological-DQD qubit states with equal joint-parity, i.e. the combined states $\ket{00}$ and $\ket{11}$ correspond to the same splitting, and $\ket{01}$ and $\ket{10}$ have the same splitting. Thus, measurement of the flux qubit splitting energy $\Delta$ is equivalent to a joint parity measurement corresponding to the projectors $\Pi_{0}$ and $\Pi_{1}$ from Eqs.~(\ref{eq:Pi_0}) and (\ref{eq:Pi_1}) acting on the topological-DQD qubit pair.
If the topological and conventional qubits are initially prepared in the superposition states $\ket{ \psi_{\rm T}}=\alpha_{\rm T}\ket{0}+\beta_{\rm T}\ket{1}$ and $\ket{\psi_{\rm C}}=\alpha_{\rm C}\ket{0}+\beta_{\rm C}\ket{1}$, respectively, then application of the even or odd parity projectors gives the (unnormalized) states
\begin{eqnarray}
\label{eq:post0}
\Pi_{0} \left( \ket{\psi_{\rm T}} \otimes \ket{\psi_{\rm C}} \right) &=& \alpha_{\rm T} \alpha_{\rm C}\ket{00} + \beta_{\rm T} \beta_{\rm C} \ket{1 1} \\
\label{eq:post1}
\Pi_{1} \left( \ket{\psi_{\rm T}} \otimes \ket{\psi_{\rm C}} \right) &=& \alpha_{\rm T} \beta_{\rm C}\ket{01} +  \beta_{\rm T} \alpha_{\rm C}\ket{10}.
\end{eqnarray}
We emphasize that the flux qubit in our proposal acts as an interferometer enabling this measurement.

{\it Topological quantum bus.}---It is now straightforward to show how one can entangle qubits and perform coherent quantum information transfer using parity measurements with the help of two flux qubits. We denote the maximally entangled Bell states (which can be used as entanglement resources) as
\begin{equation}
\left| \Phi_{\mu} \right\rangle \equiv \left( \openone \otimes \sigma_{\mu} \right) \left( \left| 01 \right\rangle - \left| 10 \right\rangle \right) / \sqrt{2}
,
\end{equation}
for $\mu=0,1,2,3$ ($\sigma_0 = \openone$). The ability to perform measurements in the Bell basis allows one to teleport quantum states~\cite{Bennett'93}, and hence transfer quantum information.

It is clear from Eqs.~(\ref{eq:post0}) and (\ref{eq:post1}) that joint parity measurements can produce entangled states, such as Bell states, but more generally we notice that
\begin{eqnarray}
\!\!\!\! \Pi_{0} &=& \left| \Phi_1 \right\rangle \left\langle \Phi_1 \right| + \left| \Phi_2 \right\rangle \left\langle \Phi_2 \right| \\
\!\!\!\! \Pi_{1} &=& \left| \Phi_0 \right\rangle \left\langle \Phi_0 \right| + \left| \Phi_3 \right\rangle \left\langle \Phi_3 \right| \\
\!\!\!\! \widetilde{\Pi}_{0} &=& \left( H \otimes H \right) \Pi_{0} \left( H \otimes H \right) = \left| \Phi_2 \right\rangle \left\langle \Phi_2 \right| + \left| \Phi_3 \right\rangle \left\langle \Phi_3 \right| \\
\!\!\!\! \widetilde{\Pi}_{1} &=& \left( H \otimes H \right) \Pi_{1} \left( H \otimes H \right) = \left| \Phi_0 \right\rangle \left\langle \Phi_0 \right| + \left| \Phi_1 \right\rangle \left\langle \Phi_1 \right|
\end{eqnarray}
where the (single-qubit) Hadamard gate is given by
\begin{equation}
H = \frac{1}{\sqrt{2}} \left[
\begin{array}{rr}
1 & 1 \\
1 & -1
\end{array}
\right]
.
\end{equation}
Hence, joint parity measurements combined with Hadamard gates fully resolves the Bell basis.

\begin{figure}
\includegraphics[width=3.0in,angle=0]{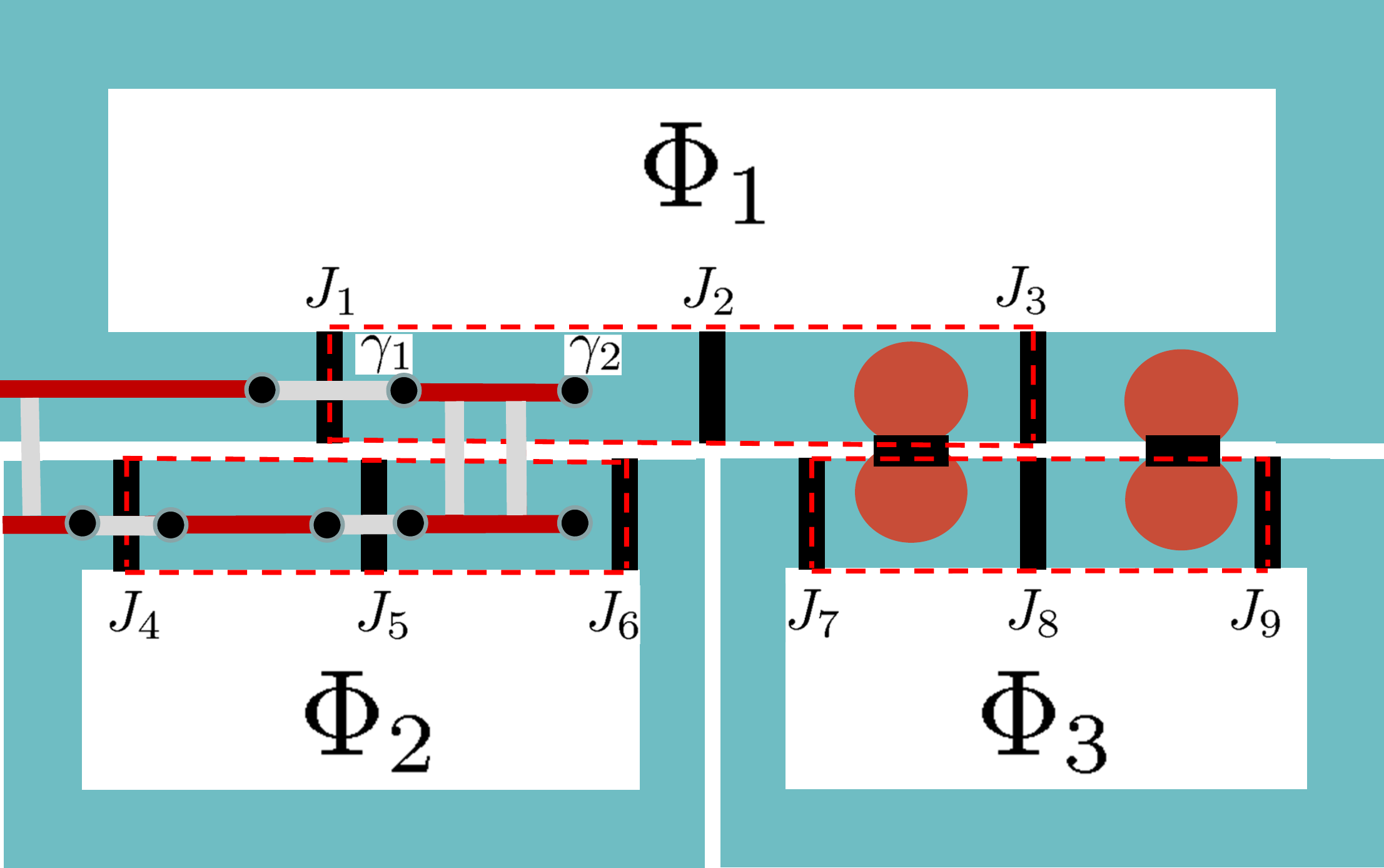}
\caption{(Color online) A proposed setup for coherent quantum information transfer between a topological and a conventional semiconductor qubit using joint parity measurements. At the degeneracy point $\Phi=h/4e$, the splitting energy of the flux qubit depends on the total charge enclosed in the region marked by the dashed (red) line. This device allows joint parity measurements of topological-conventional, topological-topological, and conventional-conventional qubit pairs, which can be used to coherently transfer information between all the different types of qubits. }
\label{Fig2}
\end{figure}

Hadamard gates can be generated (with topological-protection) by braiding Ising anyons and through standard methods for conventional qubits. As we have explained, the device in Fig.~\ref{Fig1} can be used to implement a joint parity measurement of a topological-conventional qubit pair, but it can also be used to implement joint parity measurements of topological-topological and conventional-conventional qubit pairs. Specifically, consider the setup shown in Fig.~\ref{Fig2} where there are additional flux qubits. One of these ($\# 3$) is coupled to two semiconductor DQD qubits. Again we assume that there is right-left symmetry ($E_{J_7} = E_{J_9}$) so that fluxon tunneling in the superconducting qubit allows one to measure the combined charge parity for the conventional-conventional qubit pair, as explained above. The other flux qubit ($\#2$ with $E_{J_4} = E_{J_6}$) allows one to perform joint parity measurements on topological-topological qubit pairs. The combined device allow quantum information to be transferred between topological and conventional qubits. Finally, by tuning the external fluxes $\Phi$ away from the degeneracy point one can decouple flux and conventional or topological qubits.

We conclude by remarking that the proposed joint parity measurement device not only allows one to coherently entangle and transfer information between topological and conventional systems, but also provides a new method of entangling conventional qubits, e.g. semiconductor charge or spin qubits, with each other, and hence could be useful for purely conventional systems.

{\it Note added.}---During the completion of this letter, a proposal to interface between a topological qubit in a topological insulator-superconductor heterostructure and a superconducting flux qubit via pulsed interactions that implement entangling gates was made in Ref.~\cite{Jiang10}.

\acknowledgements
We would like to thank A.~Akhmerov, S.~Das~Sarma, E.~Grosfeld, C.~Nayak, and K.~Shtengel
for enlightening discussions. We acknowledge the Aspen Center for
Physics for their hospitality.


\end{document}